\documentstyle[12pt]{article}
\begin{document}
$\;\;\;\;\;\;\;\;\;\;\;\;\;\;\;\;\;\;\;\;\;\;\;\;\;\;
 \;\;\;\;\;\;\;\;\;\;\;\;\;\;\;\;\;\;\;\;\;\;\;\;\;\;
 \;\;\;\;\;\;\;\;\;\;\;\;\;\;\;\;\;\;\;\;\;\;\;\;\;\;
 \;\;\;\;\;\;\;\;\;
  175-98-HEP $
\begin{center}
\Large
{\bf Rotating String } \\
\vskip 0.5cm
\large
Vladimir  Lugovoi

\vskip 0.1cm
{\small \em Department of High Energy Physics, Physical Technical Institute
of Tashkent , Mavlyanov st. 2b , Tashkent , \rm 700084 \small \em Uzbekistan}
\end{center}
\vskip 0.1cm
\begin{center}
{       \em Electronic addresses : lugovoi@lu.silk.org

lugovoi@physic.uzsci.net }
\end{center}

\vskip 0.7cm

\begin{abstract}
The qualitative results for string rotation in the frame of
Relativistic Flux Tube Model and the quantitative
and qualitative  results for decay
of massive open string states of string theory are used as the basic
principles in the Monte Carlo implementation for decay of massive open string.
The presented model can be used as an ingredient into any Monte Carlo model
of multiparticle production for different types of collising particles.
In the interval of total c.m. energy from 3 GeV ($e^{+}e^{-}$ annihilation) to
 1800 GeV (proton-antiproton interaction) the presented "soft" (i.e. withaut
$qq , gg , qg$ scattering ) model has agreement with experimental data
on transverse momentum distribution of secondary particles up to 4 GeV.
According to widely known theoretical hypothesis, the secondary particles
from interval $p_{T} > $1GeV are results of hard scattering QCD states.
Therefore (maybe) one can recover a question about relation between the
classical string solutions of string theory and solutions of QCD equations.
\end{abstract}

\newpage
\section{Introduction}
If we remove the quarks from QCD, there would remain a nontrivial $SU(3)$
Yang-Mills theory which must have its own spectrum of states. These states
will become the "glueball" states of QCD (See Ref.\cite{IsgurPaton}) .
The simple quark model can be
extended to incorporate such gluonic degree of freedom. In the bag
\cite{Donoghue} model, such approach is to proceed by analogy with
the "constituent quark" to posit the existence of a
"constituent gluon" with the quantum numbers
($J^{PC}=1^{--}$) of a gluon of weak-coupling perturbation theory
\cite{Jaffe}. In the non-relativistic Flux Tube model \cite{IsgurPaton}
by Isgur and Paton there is the long-range strong-coupling limit
in which the gluonic degree of freedom have condensed into collective
string like flux tubes. A flux tube is a directed element (or "string"),
and quark(antiquark) acts as a unit source of the three-dimentional flux tube.
Therefore a "junction" can annihilate (be created) there , i.e. there are
flux tube breaking and fusion. Such theoretical picture leads (see
 \cite{IsgurPaton} ) to an understanding of the linear confinement
(quark-antiquark) potential, to lineary rising Regge trajectiries
 \cite{IsgurPaton},\cite{Olsson93}. Hence, in the potential flux tube model,
a simple QCD-motivated potential \cite{Olsson93}, \cite{Eichten}
$V(r)=-\frac{k}{r}+a\cdot r$ (here $k\simeq 0.5 $ and $ a\simeq 0.2 GeV^{2}$)
is used for calculation
of interaction along to string (flux tube) lenght.

In the frame of classical analysis \cite{Olsson92} Olson, Olsson and Williams
have demonstrated that for large angular momenta the leading relativestic QCD
corrections can be interpreted as the angular momentum and angular energy of a
rotating flux tube. Therefore , the simplest interpretation of these
relativistic corrections is to consider the momentum as well as the energy
of the interacting field. So, the relativistic flux tube model
\cite{Olsson93} , \cite{OlsWil} arises as the successor to
the potential model.

In the Relativistic Flux Tube (RFT) model \cite{Olsson93}  (see Fig.1)
the energy of motion of the $i$-th quark (fermionic) is
$ H_{i}=\vec{\alpha} \cdot \vec{p_{qi}} + \beta \cdot m_{i}$. The tube is inserted by
the four momentum substitution \cite{MGOlss}
$p_{q}^{\mu} \longrightarrow p^{\mu} - p_{t}^{\mu}$ , where $p^{\mu}$ is
the new canonical four momentum and $p_{t}^{\mu}$
is the tube four momentum
computed by integration along the tube \cite{Olsson92}
\vskip 0.05cm
$\;\;\;\;\;\;\;\;\;\;\;\;\;\;\;\;\;\;(p_{ti})_{r} = 0$ ,
\begin{equation}
(p_{ti})_{\theta} = L_{ti} = \frac {r_{i}}{2 \cdot \vartheta _{\perp i}}
( H_{ti} - a \cdot r_{i} (1-\vartheta _{\perp i}^{2})^{0.5}) ,
\end{equation}
$\;\;\;\;\;\;\;\;\;\;\;\;\;\;\;\;\;\;\;\;\;\;\;\;\;\;H_{ti} = a\cdot r_{i}
( \vartheta _{\perp i}  \cdot r_{i}  \sin \vartheta _{\perp i} ) ^{-1} $,
\vskip 0.05cm
where $  \vartheta _{\perp i} $ is transverse velocity of $i$-th quark. As
one can see , the tube substitution is analogous to the introduction of the
four potential in QED. For small quark masses $m_{i} \ll ar_{i}$, it can be
shown  \cite{Olsson92} that $\vartheta _{\perp i} \rightarrow 1$
and the quarks are dynamically unimportant for large angular momenta.
In this limit we recover (see  \cite{Olsson92} ) the Nambu-Goto string
with Regge slope $\frac{L}{M^{2}} = \frac{1}{2\pi a}$.

The basic assumption of the RFT model ( see  \cite{Olsson94} ) is that the
QCD dynamical ground state for large quark separation consists of a rigid
straight tubelike color flux configuration connecting the quarks. The similar
physical picture is for classical string solutions in
Ref. \cite{Mitchell} where Mitchell, Sundborg and Turok had considered
the decay of massive open string states on the leading Regge trajectory.
These correspond to classical string solutions in the form of rigidly rotating
rods whose ends move at the speed of light (see above limit
$\vartheta _{\perp i} \rightarrow 1$ ). For these states the lenght $L$ of the
string is proportional to the mass $(L\propto M)$ .
In  \cite{Mitchell} Mitchell, Sundborg and Turok had found that the decay
rate for a string of a "classical" lenght $L$ is proportional to
$L^{(d-14)/12} \propto M^{(d-14)/12}$ , where $d$ is a space dimension.
For example, for our case $d$=4 and the decay rate
$\Gamma \propto L^{-0.83} \propto M^{-0.83} $.

Hence, rotation of string playes the fundamental role in both the flux tube
model (which has correct theoretical basis for low and middle mass
calculations) and the string theory which has theoretical description for
decay up to high mass of string (see  \cite{Mitchell} ).The question about
rotation of string has deep physical connection with a question about relation
between radial and orbital excitations, i.e. relation between momenta
$P_{r}$ and $P_{\theta} $ in Fig.1 .

As it had described in Refs.\cite{Simonov} , \cite{Olss} for
the case of large orbital
excitations $ L\gg n_{r}$ (or $L_{z}\gg n_{r}$ , where $n_{r}$ is the radial
quantum number) the system behaves as the transverse rotating string (i.e.
for the  quaziclassical long distance approximation  it is
$P_{\theta} > P_{r}$ in Fig.1 ). It is an analogy with so called
dominantly orbital excited states \cite{Olss96}.
In the opposite case  $ L\ll n_{r}$ (i.e. $P_{\theta} < P_{r}$ in Fig.1 )
the string is nearly pure inert and almost does't contribute into the kinetic
part of the Hamiltonian \cite{Simonov}.

For example, in the Lund fragmentation scheme \cite{AnderssonNilsson} ,
for the force field between quark and antiquark, an approximation
of quasiclassical relativistic massless string is
realizated\footnote{The string model becomes semiclassical when we account for
its breaking \cite{GuptaRosenzweig}.}. Hence this string does not lie on the
Regge trajectory, has not orbital momentum and rotation (i.e. the
momentum $P_{\theta}$ in Fig.1 is neglected). As it is emphasized in
 \cite{AnderssonNilsson} the surface spanned by the massless relativestic
string is always a minimal surface. This means that
(see \cite{AnderssonNilsson}), in the Lund fragmentation scheme ,
all interior properties  of massless relativistic string is determined
by the boundary.  Therefore in this approximation the
string is nearly pure inert and only (anti)quark as the excited end-point
of string has energy-momentum. Therefore, the momentum $P_{\theta}$ is
neglected and the process is one-dimensional\footnote{In the Relativistic
Flux Tube model the process is three-dimensional, because the
quark acts as a unit source of the three-dimentional flux tube (string)
which has the mass, momentum-energy and agular momentum and lies
on the Regge trajectory.}. For high radial
excitation of end-point, this approximation is correct to a high
degree of accuracy.

At the some time , after Lund string fragmentational procedure, the final
state consists of a set of string pieces \cite{AnderssonNilsson},
 which have the energies, momenta and
masses, i.e. ones lie on the linear Regge trajectories and have
rotation at rest. Hence, in Fig.1, the value $P_{\theta}$ can't be neglected.
Therefore in the Lund fragmentation model there are the classical massive
(rotating) string solutions only for small string masses (meson and baryon
hadron states).

The lund approximation is starting point for our model. According to results
of Ref. \cite{Mitchell} in our string fragmentation scheme we suppose
existence of the classical massive (rotating) string solutions not only
for small mass string states, but for the mean and large mass string
states too.

In the Relativistic Flux Tube model \cite{IsgurPaton},\cite{Olsson94}
and for classical string solutions of string theory  \cite{Mitchell},
for large quark separation a rigid straight tube like color flux
configuration connects the quarks. At the some time, in the soft
(i.e. withaut $qq , gg , qg$ scattering ) proton-(anti)proton interaction, and
in the proton-electron interaction, and in electron-positron annihilation
for high c.m. energy , the above both quark and antiquark have long
distance strong interaction , ultrarelativistic opposite momenta (in their
center of mass system) and , due to orbital momentum conservation, they move
with non-zero impact parameter. Therefore for the short time period between
formation and decay of string , there are physical conditions (for
formation and decay) which are the same as in the flux tube model or in
the classical string solutions of string theory  \cite{Mitchell}.  Hence,
at high energy interaction, the  cascade decay of massive rotating open string
can be serious realistic candidate for description of multiparticle production
by the use of cascade multistring production.

The primary string (arising in electron-positron annihilation ,
proton-electron or proton-(anti)proton interaction) has radial direction
which is directed along to summary momentum vector of both the
quark and antiquark which act as units sources of the three-dimensional
primary string (flux tube). Therefore there are correspondingly two degree
of freedom for excitation of meson (baryon) system \cite{ClosePage}
in its rest frame : the values $P_{\theta}$ and $P_{r}$ (in Fig.1).
In Refs. \cite{ClosePage},\cite{WongWangWu} it is shown that for conventional
meson states, where the flux tube is in its ground state,
the transverse confinement leads to Gauss law distribution for transverse
degree of freedom
 and the string transverse size is about a few hundred MeV (i.e.
$\sim$0.3 GeV). We use this theoretical result in our model, where the
longitudinal momentum $P_{r}$ grows with the mass of string. Therefore in
our model there is
the average summary physical picture :
\vskip 0.4cm

-the string is breaking along to direction which is determined by angle $\theta$;
\vskip 0.3cm

-the angle $\theta $ is determined by relation between the momenta
 $P_{\theta}$ and $P_{r}$

 from the previous decays; i.e. rotation of daughter string has direct

 connection with rotation of mother string
\vskip 0.3cm

-the momentum $P_{\theta}$ has "restricted" value (which is determined by

 Gauss law) , whereas $P_{r}$ grows with the string mass. Therefore the
 string

 of very large mass is breaking longitudinally (i.e. the angle $\theta$
 is about

 zero), and the string of small mass is breaking approximately
 isotropically;
\vskip 0.3cm

-the average value of angle $\theta$ grows with number of decays of string.
 This

 is natural result , which corresponds to classical (rotating) string
 solutions

 \cite{Mitchell} and to the Relativistic Flux Tube model \cite{Olsson93} where
 the string rotation

 is one in a few basic principles.
\vskip 0.3cm

As it is noted above, in Ref.\cite{Mitchell} , Mitchell, Sundborg and Turok
considered the decay of massive open string states and found that the decay
rate for a string of "classical" length $L$ (and of mass $M$)
is proportional to $L^{-0.83}$  , i.e. $\Gamma \propto M^{-0.83}$
for the space-time dimension $d$=4. The branching ratios to differ mass level
states illustrate this behaviour graphically in Fig.1 (in Ref.\cite{Mitchell}),
where the dominant decay modes are into one small string and one large string.
The same physical picture can be arise in our model , if we for simplicity
suppose\footnote{Of course, it is rough additional supposition which (in future
MC implementation ) should be changed over correct theoretical result.}
that the partial decay rate $\Gamma_{i}$ in the differential kinematical
interval of daughter strings is proportional to the
kinematical characteristics of daughter strings from this interval, i.e.
$\Gamma_{i} \propto (\varepsilon_{i}+p_{zi})^{-0.83}$ ,
where $\varepsilon_{i}$ and $p_{zi}$ are energy and $z$-projection
(i.e. along to $P_{r}$ momentum in Fig.1) of  momentum of $i$-th daughter
string in rest frame of decaying string. Therefore the (density of)
probability for decay of massive open string with mass $M$ in to
daughter strings with above kinematical characteristics
from differential interval is
$w_{i}=\frac{\Gamma_{i}}{\Gamma}\propto
 \frac{(\varepsilon_{i}+p_{zi})^{-0.83}}{M^{-0.83}}  =
(\frac{ \varepsilon_{i}+p_{zi}}{M})^{-0.83}={z_{i}}^{-0.83}$ , where
convenient variable $z_{i}=\frac{ \varepsilon_{i}+p_{zi}}{M}$ is used.
Both the $z_{1}$- and $z_{2}$-distribution functions and Gauss law
for projections of  transverse momentum (see above) of daughter string
in rest frame of decaying string allow to us to determine the masses,
energies and momentum projections of two daughter strings
in the rest frame of decaying string (see Fig.1)\footnote{According to
 Relativistic Flux Tube model , and classical
string solutions of string theory\cite{Mitchell}, for large quark
separation a rigid straight tube like color flux configuration
connects the quarks. Therefore it is necessary to stress, at
Monte Carlo simulation of decay of above rigid straight tube (string),
the $z_{1}$ and $z_{2}$ can't be generated independently.}.

The above theoretical principles and results we use in the algorithm
(see Sect.2) which is an attemption to construct the Monte Carlo
implementation  for breaking of massive open string according to
quantitative results for
decay of classical string solutions of string
theory and to the results of RFT model also.

The results of Monte Carlo calculations is discussed in Sects. 3 and 4.

\section{Decay of string}

Monte Carlo (MC) algorithm of decay of string is described in this Section.
The diagram of cascade breaking of string is shown in Fig.2.  The double
line marks the diquark.  The cascade breaking of string is described by the
iteration of an elementary process presented in Fig.3 : the string is breaking
into two non-interacting colourless daughter strings. Three schemes of decay
of the string into two strings (Fig.3) are taking into consideration
in our model. According to the Flux Tube model (see Fig.1) and in the frame
of cascade logic we generate the break of each string in its rest frame.
By every break , for example of secondary string of mass $M$
(see Fig.2) , we know following its characteristics
(in the $K_{0}$ c.m.s.\footnote{It can be, for example,
c.m.s. of $pp$-collisions, or $e^{+}e^{-}$ annihilation
c.m. system with $z$ axis along to jet axis.}) :
the scheme of the decay of secondary string, the flavours of the partons
participating in this decay, the mass $M$, the energy $E_{0}$ and the
momentum vector $\vec{P}_{0}$ of the decaying string.

    We generate the break of each string in its rest frame.
For transition from $K_{0}$ to string rest frame, let us define new
$K_{L}$ frame, where longitudinal momentum of decaying string is
equal to zero. This $K_{L}$ frame moves with
$\beta_{z0} = P_{z0} / E_{0}$ velocity along
to $z_{0}$ axis in the  $K_{0}$ frame and with Lorentz-factor
$\gamma_{0}=E_{0}/M_{\perp}=E_{0}/(E_{0}^{2}-P_{z0}^{2})^{0.5} $ , where
$M_{\perp }=(M^{2}+P_{\perp 0 }^{2})^{0.5} $ is transverse mass of $i$-th
string.

In the $K_{L}$ frame, the decaying $i-th$ string has transverse momentum
$\vec{P}_{\perp 0}$ with projections $P_{x0}$ and $P_{y0}$.
We turn the $K_{L}$ frame (in the transverse $xy$ plane) to obtain
the $x_{L}$ axis along to the vector $\vec{P}_{\perp 0}$.
The parameters of this transformation are
$ \cos\varphi_{\perp} = P_{x0}/P_{\perp 0} $ ,
$ \sin\varphi_{\perp} = P_{y0}/P_{\perp 0} $ . Let us define this new
system as $ K_{\perp} $ one , and let the $ K_{\perp}$ frame move with
$\vec{\beta}_{\perp 0} = \vec{P}_{\perp 0}/M_{\perp }$ velocity and
with Lorentz-factor $\gamma_{\perp 0}=E_{\perp 0}/M=M_{\perp }/M$ ,
where $E_{\perp 0}$ is energy of decaying string in $K_{L}$ frame.
Thereby we obtain the $K'$ rest frame of the string.

The longitudinal $z'$ axis of $K'$ frame is parallel to the longitudinal
$z_{0}$ axis of $K_{0}$ frame. But according to the Flux Tube model (see Fig.1)
the breaking string has rotation , i.e. there is nonzero angle
$\theta$ between the flux tube and the $z'$ axis in our $K'$ rest frame
(it is C.M. frame in Fig.1). At present we can't  calculate $\theta$ angle
(according to  theory and simultaneously ) in frame of our MC model ( it is
interesting problem which should be overcomed for future MC implementation ) ,
and so for continuity  of rotation  (of rigid straight tube) from the mother
to daughter string we determine the polar angle $\theta$ by the formula
$sin \theta = P_{\perp 0} / P_{1(2)}$ , where $P_{1(2)}$ is the
modulus of the momentum vector of first(second) daughter
string\footnote{The $P_{1(2)}$ value is given by eqs.(3),(20),
and so if $ P_{\perp 0} > P_{1(2)}$ , we can use an approximation
 $\sin \theta=1 $.}
in the $K'$ rest
frame\footnote{As it is emphasized in Sect.1 , in the string breaking
models \cite{AnderGustIngSjos} , \cite{AurenchBoppRanft}
the angle $\theta$ (in Fig.1) equals to zero.} , and
the azimuthal angle $\varphi$ is generated uniformly\footnote{In future
MC implementation it is necessary to take into account that the azimuthal
isotropy can be broken by angular momentum conservation. } in the interval
$(0,2\pi)$. The above formula takes into account the $\theta$
fluctuations at fixed  $ P_{\perp 0} $  and the average $\theta$
angle grows with  $ P_{\perp 0} $ of string.
In the string rest frame $K'$ the angles $\theta , \varphi $ are
determine the line of string stretch. Along this line the string break is
generated, and so we have to determine the new $K$ rest
system which is given by the $\theta , \varphi $ angles  in $K'$ system
and with the $y$  axis along to the vector product\footnote{Therefofe
third nonphysical Eulerian angle $\psi$ is absent in our calculations.}
$\vec{z} \times \vec{z'}$ .
\vskip 0.4cm

A simple model for string breaking involves quark pair creation by
the strong chromoelectric field inside the string
(see \cite{GuptaRosenzweig}) . The flavours of quarks, produced from vacuum
at the decay of string, are generated according to the
relation  \cite{AnderGustIngSjos}
\begin{equation}
 u \;\; : \;\; d \;\; : \;\; s \;\; = \;\; 3 \;\; : \;\; 3 \;\; : \;\; 1 \;\;\;\;.
\end{equation}

In the RFT model  \cite{ClosePage},\cite{WongWangWu}
the transverse confinement leads to Gauss law distribution for transverse
degree of freedom  and the string transverse size is about
a few hundred MeV. We use this theoretical result in our model, where
in the $K$ frame $P_{x1}$ and $P_{y1}$ components of
the transverse momentum $P_{\perp}$
of first daughter string are generated according to Gauss law  (see Sect.1)
\begin{equation}
r_{1}(r_{2}) = \sigma^{-1} (2\pi)^{-0.5}
\int_{-\infty}^{P_{x \; 1}(P_{y \; 1})} exp(-x^{2}/2\sigma^{2})
\;\;\; dx  \;\;\;\;\;\;  ,
\end{equation}
where $r_{i}$ is uniformly distributed random number in the range (0,1)
and $\sigma$ is parameter, which is $\sigma_{qq}$ = 0.35 GeV for
the string decays in Figs.3b and 3c. For the string decay in Figs.3a
the model parameter $\sigma$ is $\sigma_{u(d)}$=0.25 GeV
(if valence quarks are u(d)-quarks), and the model parameter $\sigma$ is
$\sigma_{s}$=0.30 GeV (if only one valence quark is s-quark).

    In the $K$ frame we determine (see Sect.1) the following variables:
\vskip 0.5 cm
$z^{+}_{1} = (E_{1}+P_{z1})/M  , \;\;   z^{+}_{2} =(E_{2}+P_{z2})/M  , \;\;
z^{-}_{1} = (E_{1}-P_{z1})/M  , \;\; $
\begin{equation}
                                          z^{-}_{2} =(E_{2}-P_{z2})/M  ,
\end{equation}
where $E_{1}$ and $E_{2}$ ($P_{z1}$ and $P_{z2}$) are energies
( the momentum projections
(on $z$ axis)) of first and second daughter strings. Generally speaking,
the $z$-distribution can be  singular one . In this case the $z$-variable
should be generated from interval ( $z_{min} , z_{max}$ ). The possible choice for
$z_{min}$ and  $z_{max}$ is the following.  According to law of the
conservation of energy, at the production of first daughter
string we generate the variable  $z^{+}_{1}$  in the interval
\begin{equation}
  z^{+}_{1min} \;\;\; < \;\;\; z^{+}_{1} \;\;\; < \;\;\; z^{+}_{1max}  \;\;\;   ,
\end{equation}
where
\begin{equation}
  z^{+}_{1min} \;\;\; = \;\;\;(E_{1} \;\;\; -
\;\;\; \mid P^{max}_{z1} \mid )/M  \;\;\;  ,
\end{equation}
\begin{equation}
  z^{+}_{1max} \;\;\; = \;\;\;(E_{1} \;\;\; +
\;\;\; \mid P^{max}_{z1} \mid )/M  \;\;\;  ,
\end{equation}
where
\begin{equation}
  E_{1} \;\;\; = \;\;\;(M^{2} \;\;\; +
\;\;\; (M^{min}_{1})^{2} \;\;\; - \;\;\; (M^{min}_{2})^{2})/(2M) \;\;\; ,
\end{equation}
\begin{equation}
  \mid P^{max}_{z1} \mid  \;\;\; = \;\;\; ((E_{1})^{2} -
(M^{min}_{1})^{2} )^{0.5} \;\;\; ,
\end{equation}
where  $M^{min}_{1}$  and  $M^{min}_{2}$  are
the minimal masses of first and second
daughter strings (see Fig.3). Any daughter string is converted into a
hadron if it cannot be broken into two lightest hadrons , i.e.
the minimal mass of first (second) daughter string equals to the
sum of the masses of two lightest hadrons, which have the
flavours depending on the quark  contents of first (second)
daughter string. Therefore (for simplification) the minimal mass of
first (second) daughter string can be determined as the double
mass of lightest hadron with quark composition of first (second)
daughter string (see Appendix 1).

\vskip 0.4cm

     For production of first daughter string, the
distribution density of $z^{+}_{1}$ variable (for all schemes of the
string breaking in Fig.3) is parameterized by the form (see Sect.1)
\begin{equation}
           f(z) \;\; = \;\; z^{-0.83} \;\;  .
\end{equation}
The same scaling\footnote{According to theory of decay
of massive open string \cite{Mitchell},
in our Monte Carlo implementation the distribution of $z$ does not depend on
the mass of mother (decaying) string, i.e. this distribution is scaling one.}
function (10) can be used for production of second
daughter string if we generate the variable $z^{-}_{2}$ (see eq.(3)). In the
$K$ frame the equalities (4) lead to ones :
\begin{equation}
        z^{+}_{1} \;\;  + \;\; z^{+}_{2} \;\; = \;\; 1  \;\; ,
\end{equation}
\begin{equation}
        z^{+}_{1} \;\; . \;\; z^{-}_{1} \;\; =
\;\; ((M_{1})^{2} \;\; + \;\; (P_{\perp})^{2} )/M^{2}  \;\;  ,
\end{equation}
\begin{equation}
        z^{+}_{2} \;\; . \;\; z^{-}_{2} \;\; =
\;\; ((M_{2})^{2} \;\; + \;\; (P_{\perp})^{2} )/M^{2}  \;\;  ,
\end{equation}
where $M_{1}$ and $M_{2}$ are the masses of first and second daughter
strings (see Fig.2).Therefore (after generation of the
variable $z^{+}_{1}$ ) the variable  $z^{-}_{2}$
is generated in the $K$ frame (according to the scaling
function (10)) in the interval
\begin{equation}
  z^{-}_{2min} \;\; < \;\; z^{-}_{2} \;\; < \;\; z^{-}_{2max} \;\; ,
\end{equation}
where
\begin{equation}
  z^{-}_{2min} \;\; = \;\; ((M^{min}_{2})^{2} \;\; +
\;\; (P_{\perp})^{2}) \;\; / \;\; (M^{2}(1-z^{+}_{1})) \;\; ,
\end{equation}
\begin{equation}
  z^{-}_{2max} \;\; = \;\; 1 \;\; - \;\;
((M^{min}_{1})^{2}+(P_{\perp})^{2})/(M^{2} z^{+}_{1}) \;\;
\end{equation}
are determinated according to law of the conservation of energy.
In the $K$ frame from the conservation of energy and momentum after
generation of  $z^{+}_{1}$  and $z^{-}_{2}$  we can find (see (4)) the
variables
\begin{equation}
                       z^{-}_{1} \;\;  =
\;\; 1 \;\; - \;\; z^{-}_{2} \;\;    ,
\end{equation}
\begin{equation}
                       z^{+}_{2} \;\;  =
 \;\; 1 \;\; - \;\; z^{+}_{1} \;\;   .
\end{equation}
The energies and longitudinal momenta of daughter strings are
 given by formulas (see Appendix 2)
\begin{equation}
           E_{i} \;\;  = \;\;  M( z^{+}_{i} \;\; +
\;\; z^{-}_{i})/2 \;\; ,
\end{equation}
\begin{equation}
           P_{zi} \;\; = \;\; M( z^{+}_{i} \;\; -
\;\; z^{-}_{i})/2 \;\; , \;\;      i=1,2  .
\end{equation}

Because of an equality  $z^{+}_{1} =(1 - z^{+}_{2})$ and  from
eq.(10) one can see that
the variable $z^{+}$ has $z^{-0.83}$  distribution for
first daughter string and
(according to energy conservation) $(1 - z)^{-0.83}$ distribution for
second daughter string. Therefore (in the frame of
presented model) for the
fitted values of the free parameters,
for small mass of mother string and
for the random numeration of daughter strings,  at the decay of mother
string into two daughter strings the distributions of first
and second daughter
strings on the part of energy(and momentum) of mother string
are similar to momentum distributions for valence quark and
antiquark in meson\footnote{
For $z^{-}$ variable there is the same result because the $z^{+}_{1}$ and
$z^{-}_{2}$ variables have the same parametrization (10).},
i.e. similar to  $f(z) = z^{-0.5} (1-z)^{-0.5}$
(see Ref. \cite{Innocente} ).

    In the models \cite{AnderGustIngSjos},\cite{AurenchBoppRanft}
there is no such result, because
for the force field between quark and antiquark, an approximation
of quasiclassical relativistic massless string is
realizated (see \cite{AnderssonNilsson} and Sect.1)
, therefore the scaling functions are used for the other
process (see Fig.4a and Sect.1). For example,
in Ref.\cite{AurenchBoppRanft} the scaling function
for conversion of the quark into
meson+quark is equal to $f(z) = 1-a+3a(1-z)^{2}$ , with $a$=0.88 , in
Ref.\cite{AnderGustIngSjos} $f(z) = (1+c)(1-z)^{c}$ ,
with  $c \approx  0.3 \div 0.5$.

    After determination of the momenta of daughter strings, we control
the inequalities
\begin{equation}
         E_{i} \;\;\; > \;\;\; P_{i} \;\;\;   ,   \;\;\;     i=1,2
\end{equation}
If inequalities (21) are fulfilled for the i-th string, we determine the
string mass $M_{i}=((E_{i})^{2} -(P_{i})^{2})^{0.5}$
and the decay scheme of $i$-th string
and flavours of quarks, produced in this decay from vacuum, are
generated. Then we control the inequality
\begin{equation}
M_{i} \;\; > \;\; M^{min}_{i1} \;\; + \;\; M^{min}_{i2} \;\;  ,
\end{equation}
where the $M^{min}_{ij}$ is minimum mass of the $j$-th hadron with the fixed
quark composition, which can be produced by decay of $i$-th string.
The decay scheme (see Fig.3) is generated according to
the relation \cite{AnderGustIngSjos}
\begin{equation}
       a  \;\; : \;\; b \;\; = \;\;  0.95 \;\; : \;\; 0.05 \;\;\;\;.
\end{equation}
A model has six parameters ( see eqs. (2),(3),(10),(23) ),
which have correct theoretical validity.

      If the inequalities (21),(22) are fulfilled for both daughter strings,
their decays are simulated. The algorithm for decay of daughter
string of the mass $M_{1} (M_{2})$ is similar to algorithm for decay
of secondary  string of the mass $M$ .

     If the inequalities (21),(22) are fulfilled for only one
daughter string (for definiteness we take the first string),
the decay of the
mother string into first daughter string and hadron is generated. If
$E_{2}\;\; < \;\; P_{2}$ we select the lightest
hadron from hadrons with the given quark composition. The early generated
$z^{+}_{1}$ , $P_{x1}$ and $P_{y1}$ values are not changed, but we generate
the hadron mass $M_{2}$ instead of $z^{-}_{2}$ , if  the hadron is resonance.
After determination of $z^{+}_{2}$ from (18), we determine
$z^{-}_{2}$ according to (see (13))
\begin{equation}
  z^{-}_{2} \;\; = \;\; ((M_{2})^{2} \;\;   + \;\; (P_{\perp})^{2}
) \;\; / \;\; (M^{2} z^{+}_{2})  \;\;  ,
\end{equation}
and  $z^{-}_{1}$ from (17). The energies and longitudinal momenta of the
daughter string and hadron are calculated from (19),(20) (see Appendix 2).
If the inequalities (21),(22) are fulfilled for the daughter string, the
algorithm of it decay is similar to algorithm for decay of secondary
string of the mass $M$ .

      The decay of mother string into two hadrons is generated, if at the
decay of the mother string into two strings the inequalities (21),(22)
 are not fulfilled for the both strings, or if at the decay of mother string
into a string and a hadron the inequalities (21),(22) are not fulfilled
for the daughter string. There are some decay modes for the given quark
content of these hadrons. We attribute a weight to each decay mode. This
 weight is equal to the product of three factors. The spin factor is equal
to $(2J_{1}+1)(2J_{2}+1)$, where $J_{1}$ and $J_{2}$
are spins of the hadrons. The kinematic
 factor is equal to the two body phase space volume or to zero, if the
string mass $M$ is smaller than the sum of the masses of daughter hadrons.
 The $SU_{3}$-factor is taken into consideration, if there are several
hadrons with the same quark content, spin and parity. For example,
$SU_{3}$-factor of $\eta$-meson which is formed
from $\overline{u}u$-pair is equal
to $1/6$ , and the same $SU_{3}$-factor of $w$-meson is equal to $1/2$.

 The $M_{1}$ and $M_{2}$ masses of resonances are generated after
the generation of the decay mode. For example, if string decays
into two resonances and the mass of first resonance must be
generated at first, the $M_{1}$ and $M_{2}$ masses can be generated
according to the Breit-Wigner distributions in intervals
\begin{equation}
  M^{min}_{1} \;\;  <  \;\; M_{1} \;\; < \;\; M - M^{min}_{2} \;\; ,
\end{equation}
\begin{equation}
  M^{min}_{2} \;\;  <  \;\; M_{2} \;\; <  \;\; M - M_{1} \;\; ,
\end{equation}
where $M^{min}_{i}$ is maximum sum of masses of particles produced
by the decay of $i$-th resonance. In the $K$ frame the square of
transverse momentum of resonance $P_{\perp}^{2}=P_{x1}^{2} \; + \; P_{y1}^{2}$
is generated in the interval $(0,P^{2})$, and the azimuthal
angle is generated uniformly in the interval $(0,2\pi)$. The momentum of
first resonance is supposed to have the sharp angle with z axis.
 The decays of unstable hadrons into stable and quasistable
particles are generated in the $K$ frame, for example, according
to algorithm  \cite{ChudLu} .

     The momenta of stable and quasistable particles are transformed from
the $K$ frame to the $K'$ frame according to formulas
\begin{equation}
p'_{xi} = - p_{xi} \; cos\theta \; cos\varphi \;  -
\; p_{yi} \; sin\varphi \;-\; p_{zi} \; sin\theta \; cos\varphi   \;\; ,
\end{equation}
\begin{equation}
p'_{zi} =   p_{xi} \; sin\theta \; - \; p_{zi} \;
cos\theta   \;\;\;\;\; ,
\end{equation}
\begin{equation}
 p'_{yi} = - p_{xi} \; cos\theta sin\varphi  +
p_{yi} \; cos\varphi \; - \; p_{zi} \;
sin\theta \; sin\varphi  \;\;\;  ,
\end{equation}
where $p_{xi}$ , $p_{yi}$ , $p_{zi}$ are the
momentum projections of $i$-th
particle (in the $K$ frame), and $ \theta , \varphi $ angles are
determined  above.
After Lorentz transformation of the energies and momentum vectors of particles
from the $K'$ frame to the $K_{\perp}$ frame,
momentum vectors of particles are transformed from the $K_{\perp}$
to $K_{L}$ frame according to the parameters of transformation
$ \cos\varphi_{\perp}$ , $ \sin\varphi_{\perp} $ (see above) in the
transverse  $xy$ plane.  After Lorentz transformation of the
energies and momentum vectors of stable and quasistable particles
from the $K_{L}$  frame to the $K_{0}$ frame we know all characteristics of
secondary particles in the $K_{0}$  c.m.s.
\vskip 0.5 cm

\section{$e^{+}e^{-}$ annihilation}
\vskip 0.1 cm
{\bf3.1 $e^{+}e^{-}$ annihilation at low energy }
\vskip 0.5 cm

In Sect.2 a cascade model of string breaking (see Figs.2,3)
had been constructed as a new model for hadronization.
In $e^{+}e^{-}$ annihilation with center of
mass energy $E_{c.m.}$ up to a few GeV the process of gluon emission
\cite{Gottshalk} -  \cite{TSjo}
practically is absent
\cite{Gottshalk} -
 \cite{Hanson} .
Therefore we deal
with only one primary quark-antiquark string (see Fig.5). Therefore in
this energy interval there is unique possibility to check MC cascade model
of string breaking (Sect.2). However, our model
incorporates the production and decay the hadrons which are
constructed by $u$, $d$, $s$ quarks only. Therefore there is a possibility to
check this model at center of mass energy only which is under threshold
of charm quark pair production, i.e. at c.m.energy equals to 3 GeV
\cite{Siegrist}, \cite{Hanson} .
\vskip 0.5 cm

{\bf3.2  The formation of the primary string and its decay}
\vskip 0.5 cm

   The $e^{+}$ and $e^{-}$ annihilate to
form a virtual photon which produces a
quark-antiquark pair (Fig.5). In the presented model the flavour of the
quark $q$ (and antiquark $\overline{q}$ ) (Fig.5)
is generated according to relation (between the probabilities for
$u\overline{u}$ , $d\overline{d}$ ,
$s\overline{s}$  quark-antiquark pair
production)
\begin{equation}
                       u:d:s = 4:1:1     \;\;\;    .
\end{equation}
In the process $e^{+}e^{-}\rightarrow q\overline{q}$ (in
the  $e^{+}e^{-}$ center mass system) the angle
distribution for the quark is given by the form
\begin{equation}
   N^{-1} dN/d \Omega = (1+cos^{2}\theta) \;\; 3/16\pi  \;\;\; .
\end{equation}
where $\theta$ is polar angle between the
quark momentum vector and electron
momentum vector, $d\Omega = 2\pi sin\theta d\theta$ .

       A quark $q$ and antiquark $\overline{q}$
(Fig.5) stretch the primary string $A$
which decays into secondary hadrons
according to the algorithm of cascade of
string breaking which is described in
detailes in Sect.2. In $e^{+}e^{-}$  c.m.s.
the quark momentum vector determinates the direction of the string
breaking (see Sects. 1,2).
\vskip 0.5 cm

{\bf3.3 Comparison with experimental data for $e^{+}e^{-}$ annihilation }
\vskip 0.5 cm

In $e^{+}e^{-}$ center of mass system (see Fig.5) the quark momentum vector
is opposite to momentum vector of antiquark. Therefore after hadronization
there are two jets of secondary hadrons . A direction for jet axis (see
Fig.6) is determinated by Monte Carlo method
( \cite{Siegrist}, \cite{Hanson} ) .

At $E_{c.m.}$= 3 GeV the time for Monte Carlo simulation (see above) of
one $e^{+}e^{-}$ annihilation is 0.40 second. Illustration of a hadronic event
from $e^{+}e^{-}$ annihilation is in Fig.6. The results of calculations
according to the model are the curves in Fig.7
(the multiplicity distribution), Fig.8 (rapidity distribution),
Figs.9 ($p_{\perp}$-distribution). The
variables $x$ and $x_{\parallel}$ (see Figs.8,9) are defined
by  $x=2P/E_{c.m.}$ and $x_{\parallel}=2P_{\parallel}/E_{c.m.}$
(see Fig.6 and Refs. \cite{Siegrist}, \cite{Hanson} ).
\vskip 0.5 cm

\section{Conclusion}

The last few years have witnessed rapid progress in the description of
string-string interactions \cite{Werner} and semiclassical decays
of strings (flux tubes). For example, in Ref.\cite{GuptaRosenzweig}
Gupta and Rosenzweig explored
the implications of string breaking (flux tube fission) for hadron decays ;
and in Ref.\cite{Nussinov} Nussinov shown that the flux tube model can be
successfully applied to analytical calculations for various stages of
chromoelectric flux tube intersection, rearrangement, and eventual
fragmentation ("hadronization") of the resulting highly excited flux tubes ;
in the Ref.\cite{Mitchell} Mitchell { \em et. all } calculated the decay
rates for arbitrarily massive states on the leading Regge
trajectory for open strings.

In our Monte Carlo (MC) model \cite{VLu} we attempt
to construct the string (and its decay)
which has the physical properties the same as the string in the Relativistic
Flux Tube model  \cite{Olsson93} , \cite{OlsWil} , \cite{Olsson92} and
simultaneously the same as the classical massive open string solutions
\cite{Mitchell} of string theory. Therefore in our MC model the string decay
is computed taking the effect of string rotation into account, it is the first.
The second, in the Ref. \cite{Mitchell} Mitchell, Sundborg and Turok have
calculated the decay of the fundamental string in an arbitrary number of
dimensions. For calculation of decay probability we use
decay rates \cite{Mitchell} for the space-time dimension $d$=4.

We can make a comparison with some results on fundamental (Nambu-Goto)
strings and RFT model. In the decay of a fundamental string
\cite{Mitchell} in the critical dimention $d$=26
Mitchell, Sundborg and Turok find that the string is likely to break
symmetrically (in contrast to the noncritical dimention) into two states,
each with roughly half the (mass) excitation number of the original state.
The same result is for semiclassical decay of excited string states
in the flux tube model \cite{GuptaRosenzweig} , where
the space-time dimension $d$=4.
At the same time
at the decay of a fundamental string
\cite{Mitchell} in noncritical dimension  the string decays by emitting a
state close to the ground state and a highly excited state.
In our Monte Carlo model we use
decay rates (from Ref. \cite{Mitchell}) for the space-time dimension $d$=4.
Hovewer, because of momentum-energy conservation we generate the kinematical
characteristics of the both daughter strings from intervals
which are determined by the conservation laws, the first. The second,
because of isotropy of time-space, and because of
$P$ (space inversion) invariance of strong interaction,
we can suppose that in the
statistical set of decays of mother string (with fixed mass)
the distributions for kinematical characteristics for first daughter
string are the same as ones for second daughter
string\footnote{At the random numeration of both strings.} inside
of the above kinematical intervals\footnote{See choice for intervals and
distributions for simulation of $z_{1}$ and $z_{2}$ variables from Sect.2.}.
Therefore our MC calculations yield result that the share  of
 string decays into the state close to the hadron state
 and a highly excited state is about 25$\%$ only.

The some other properties of our model one can see by comparison with
electron-positron experimental data at the total center of mass energy
equal to 3 $GeV$ , where the process of gluon emission
practically is absent and  we can
check model at center of mass energy which is under threshold
of charm quark pair production\footnote{In our hadronization scheme
only lightest $u$- , $d$- , $s$-quarks is used.}.

The most interesting quantitative results for proton-(anti)proton interactions
will be given elsewhere\footnote{The
preliminary results see, for example, in \cite{Lug}.}, but now it is necessary to
stress one interesting property which has direct connection with massive
open string dynamics \cite{Mitchell}.

Because of the rotation (Fig.1) of decaying string (Fig.3) , the average
transverse momentum of secondary strings (Fig.2) grows with the mass of
primary string ( the A string in Figs.2,5). A quantitatively illustration of
this is in Fig.10. In other side the average mass of primary string grows
with the total energy of collising particles ( $e^{+}e^{-}$ ,
$\overline{p}p$ , $pp$ ) . Therefore the average transverse momentum of
secondary strings grows with the total energy of collising particles, and so
for the interval of total energy (in the center of mass system of collising
particles) from 3 GeV ($e^{+}e^{-}$ annihilation) to 1800 GeV
(proton-(anti)proton interactions\footnote{In proton-(anti)proton interactions
for formation of primary strings, the well-known soft version of dual parton
model \cite{Capella} is used, i.e. hard interactions ($qq$, $gg$, $qg$) between
partons of collising hadrons is not calculated.}) there is agreement between
experimental data and our theoretical MC calculations on transverse momentum
$p_{\perp}$ distribution of secondary particles up to $p_{T} = 4 GeV$
 (see Figs.11-13) . It is
interesting because there is widely known theoretical hypothesis that in the
$p_{\perp}$ interval (approximately) from $1 GeV$ to $4 GeV$ the particle
production is a result of QCD hard ($qq$, $gg$, $qg$) scattering states
(see \cite{Werner}). But in our model, all secondary particles are result only
of decay \cite{Mitchell} of states of massive open strings (classical string
solutions of string theory \cite{Mitchell} ) and with the same behaviour as for
particles which are result of QCD hard scattering states\footnote{The detailes
 for production of secondary particles which are the result of QCD hard
($qq$, $gg$, $qg$) scattering states , and the detail discussion about
$p_{\perp}$ distribution are in Refs. \cite{Andersson} , \cite{Ranft} ,
\cite{Sjostrand}.} (at least in the $p_{\perp}$ scale).
Therefore (maybe) one can recover a question about relation between the
classical string solutions of string theory and solutions of QCD equations.

Our results are a hint that it would be very interesting to construct
the Monte Carlo implementation for the decay of massive open string most
closely to results of string theory\footnote{
For example, for MC implementation it is best to have the strict
(without our additional suppositions)
theoretical partial decay rate , which depends on the kinematical
characteristics of daughter string, it is the first.
The second, for MC
scheme for long-distance interaction between quark and antiquark,
it it necessary to have a theoretical (share or probability for)
relation between the states with angular (orbital) momentum
$ L = 0$ (here the quark degrees  of freedom dominate) and the
states with angular (orbital) momentum $ L \gg 1$ (here the string degrees
of freedom dominate).}.
We hope that our model is first step to
this aim. But already the presented string breaking model can be used as an
ingredient into any Monte Carlo model of multiparticle production
for different types of collising particles ( $\overline{p}p$ , $e^{+}e^{-}$ ,
$e^{-}p$ , heavy-ion reactions ets).

For heavy-ion reactions there is widely known theoretical prediction
that hard parton-parton interaction and formation of quark-gluon plasma
are two sources of secondary particles with high transverse momenta.
We hope that our calculations shown that the string rotation is also
the source of high transverse momenta, and so this result can't be ignored
in the realistic calculations of high energy reactions.

\vskip 0.8cm

\begin{center}
ACKNOWLEDGMENTS
\end{center}
The author would like to express his sincere thanks to
V.M.Chudakov for many discussions.

\vskip 0.5 cm

{\large  \bf Appendix 1}
\vskip 0.5 cm

According to experimental data \cite{AlnerGJ} for high
energy hadron-hadron collisions,
for the produced primary hadron from every hadron multiplet
the share of the members with  big mass grows with total energy of collision.
Therefore there is second possibility for determination of the minimal
mass (of first (second) daughter string), i.e. in second case
it is the product of two factors.
The first factor equals to 2. The second factor
grows with energy. For example, at  $\sqrt{s} <$ 100 GeV  the second
factor is the mass
of lightest hadron with quark composition of first (second) daughter
string; at 100 GeV $< \sqrt{s} <$ 1000 GeV  the second factor is (with equal
probability) the mass of any hadron with quark composition of first
(second) daughter string; at  $\sqrt{s} >$ 1000 GeV  the second factor is the
bigest mass of the hadron with quark composition of first (second)
daughter string. This determination of the minimal mass
of the daughter strings leads to the better agreement with
the experimental data, but it is not a decisive factor.
\vskip 0.5 cm

{\large  \bf Appendix 2}
\vskip 0.5 cm

At decay of string of the big mass the sharp angle $\theta$ (in Fig.1)
leads to the sharp angle between the $z$
axis and the momentum vector of the daughter string.
Therefore it is necessary to rewrite the formulas (19)
as the following
\vskip 0.5 cm
$P_{z1} =   M \mid z^{+}_{1} - z^{-}_{1} \mid /2   \;\;\; $  ,
$P_{z2} = - M \mid z^{+}_{2} - z^{-}_{2} \mid /2   \;\;\; $  .

\newpage
\begin{center}
{\large  \bf Figure Captions}
\end{center}
\vskip 0.7 cm

{\large \bf Fig.1 } In the Flux Tube model \cite{Olsson93} : the portion of a
             meson consisting of a segment of flux tube from the
             center of momentum to the $\em i^{th}$ quark.
\vskip 0.8 cm
{\large \bf Fig.2 } The diagram of our model for the cascade break of
             the primary string $A$ . $\overline{1}$ and $2$ are
             the partons stretching the secondary string of mass  $M$ .
             $M_{1}$ and $M_{2}$ are the masses of daughter  strings
             produced at the break of the mother string. $\overline{1}$ is
             antiquark, $2$ is quark.
             Here and below the double line is diquark.
\vskip 0.8 cm
{\large \bf Fig.3 } Shemes of the string breaking in our model.
\vskip 0.8 cm
{\large \bf Fig.4 } {\bf a)} An elementary Monte Carlo process
             of the Models \cite{AnderGustIngSjos} , \cite{AurenchBoppRanft} :
              $q \rightarrow q'+ h$ .
             {\bf b)} The diagram of the Lund model \cite{AnderGustIngSjos}
             (and model \cite{AurenchBoppRanft})
             of breaking of the colour singlet $q'\overline{q}'$ string into
             hadrons $h$ and secondary small mass $q''\overline{q}''$
             string , which is been brouken into two hadrons.
\vskip 0.8 cm
{\large \bf Fig.5 }
         At the low energy $E_{c.m.}$ by $e^{+}e^{-}$ annihilation
         a single virtual photon $\gamma$  produces a quark-antiquark
         ($q\overline{q}$) pair, tensing the primary string $A$ ,
         which decays (see Fig.2) into secondary hadrons according to
         algorithm of cascade model of string breaking.
\vskip 0.8 cm
{\large\bf Fig.6}
         Illustration \cite{Hanson} of a hadronic event from $e^{+}e^{-}$
         annihilation showing the jet axis and the components of the momentum
         $\vec{p}$ of a particle parallel to ( $p_{\parallel}$ ) and
         perpendicular to ( $p_{\perp}$ ) the jet axis.
\vskip 0.8 cm
{\large\bf Fig.7} Charged particle multiplicity distribution.
         $K^{0}_{s} \rightarrow \pi^{+} \pi^{-}$ decays are included
         \cite{Siegrist} , \cite{Hanson}.
         Here and below the line is the calculation according to the model.
\vskip 0.8 cm
{\large\bf Fig.8} Particle-density distribution $\sigma^{-1}d\sigma / dy$
         vs $y$ for jets (see Fig.6) with $x_{max} > 0.3$.
         $x_{max}$ is the highest-$x$ particle on one side of the event.
         The jet direction is oriented so that $x_{max}$ is at positive $y$.
         $y$ is the rapidity of the particle relative to the jet direction
         assuming a pion mass. $x_{max}$ is at positive $y$ and
         is not plotted. The distributions are normalized to the cross
         sections for jets with $x_{max} > 0.3$.
         Experimental data are from  \cite{Siegrist} , \cite{Hanson}.
\vskip 0.8 cm
{\large\bf Fig.9}  Particle-density distribution
         $\sigma^{-1} d\sigma / dp_{T}$ vs $p_{T}$ for particles
         opposite ( negative $x_{\parallel}$ ) jets with $x_{max} > 0.3$.
         $p_{T}$ is the component of particle momentum perpendicular
         to the jet direction (see Fig.6).
         Experimental data are from  \cite{Siegrist} , \cite{Hanson}.
\vskip 0.8 cm
{\large\bf Fig.10.} The theoretical MC dependence of the
        average transverse momentum of secondary strings on
        the mass of primary string A (see Figs.2,5 ).
\vskip 0.8 cm
{\large\bf Fig.11.} The dependence of the inclusive production cross section of
        charged particles on the transverse momentum $p_{T}$ . Experimental
        data are from \cite{Albajar}.
\vskip 0.8 cm
{\large\bf Fig.12.} The same as in Fig.11 .
         Experimental data are from \cite{Albajar}.
\vskip 0.8 cm
{\large\bf Fig.13.} The same as in Fig.11 .
         Experimental data are from \cite{Abe}.

\end{document}